\documentclass[12pt]{article}
 \pdfoutput=1
\usepackage{amsfonts,amsthm,amsmath,amssymb,slashed}
\usepackage[textwidth = 500 pt, textheight = 630 pt]{geometry}
\usepackage{latexsym,amsfonts,amsmath,amssymb,theorem,dsfont,wasysym}
\usepackage{amssymb}
\usepackage{amsmath}
\usepackage{amstext}
\usepackage{graphicx,epsfig}
\usepackage{epsfig}
\usepackage{verbatim} 
\usepackage{fancyhdr}
\usepackage{fancybox}
\usepackage{color}
\usepackage{ulem,bbold}
\usepackage{enumitem}
\usepackage{subfigure}
\usepackage{bbm}
\usepackage{parskip}
\usepackage[numbers,sort&compress]{natbib}

\definecolor{MyDarkBlue}{rgb}{0.15,0.15,0.45}
\usepackage[linktocpage=true]{hyperref}
\hypersetup{
colorlinks=true,
citecolor=MyDarkBlue,
linkcolor=MyDarkBlue,
urlcolor=MyDarkBlue,
}

\usepackage[numbers,sort&compress]{natbib}
\usepackage{hypernat}

\def\beq{\begin{eqnarray}}
\def\eeq{\end{eqnarray}}

\def\({\left(}
\def\){\right)}

\def\f{\varphi}
\def\hh{\hat h}

\newcommand{\be}{\begin{equation}}
\newcommand{\ee}{\end{equation}}
\newcommand{\la}{\langle}
\newcommand{\ra}{\rangle}
\def\ea{\end{eqnarray}}
\def\ba{\begin{eqnarray}}

\def\beq{\begin{eqnarray}}
\def\eeq{\end{eqnarray}}

\def\({\left(}
\def\){\right)}
\def\mn{_{\mu \nu}}

\def\p{\partial}
\def\la{\langle}
\def\ra{\rangle}

\def\lsim{\mathrel{\rlap{\lower3pt\hbox{\hskip0pt$\sim$}}
     \raise1pt\hbox{$<$}}}         
\def\gsim{\mathrel{\rlap{\lower4pt\hbox{\hskip1pt$\sim$}}
     \raise1pt\hbox{$>$}}}         

\def\lsim{\mathrel{\rlap{\lower3pt\hbox{\hskip0pt$\sim$}}
     \raise1pt\hbox{$<$}}}         
\def\gsim{\mathrel{\rlap{\lower4pt\hbox{\hskip1pt$\sim$}}
     \raise1pt\hbox{$>$}}}         

\usepackage{amsmath}
\usepackage{amsfonts}
\usepackage{verbatim}
\usepackage{graphicx}
\usepackage{subfigure}
\usepackage{amssymb}
\usepackage[T1]{fontenc}
\usepackage{slashed}

\begin{document}

\renewcommand{\thefootnote}{\fnsymbol{footnote}}

\makeatletter
\@addtoreset{equation}{section}
\makeatother
\renewcommand{\theequation}{\thesection.\arabic{equation}}

\rightline{}
\rightline{}
   \vspace{1.8truecm}


\vspace{10pt}


\begin{center}
{\Large \bf{Slavnov-Taylor Identities for Primordial Perturbations}}
\end{center} 
 \vspace{1truecm}
\thispagestyle{empty} \centerline{{\large  {Lasha  Berezhiani and Justin Khoury}}
}

\vspace{1cm}
\centerline{{\it Center for Particle Cosmology, Department of Physics \& Astronomy, University of Pennsylvania,}}
 \centerline{{\it  209 South 33rd Street, Philadelphia, PA 19104}}
 
 \vspace{1cm}
 
\begin{abstract}
Correlation functions of adiabatic modes in cosmology are constrained by an infinite number of consistency relations, which relate
$N+1$-point correlation functions with a soft-momentum scalar or tensor mode to a symmetry transformation on $N$-point correlation
functions of hard-momentum modes. They constrain, at each order $n$,  the $q^n$ behavior of the soft limits. In this paper we show that all consistency relations
derive from a single, master identity, which follows from the Slavnov-Taylor identity for spatial diffeomorphisms. This master identity is valid at any value of $q$
and therefore goes beyond the soft limit. By differentiating it $n$ times with respect to the soft momentum, we recover the consistency relations
at each $q$ order. Our approach underscores the role of spatial diffeomorphism invariance at the root of cosmological consistency relations. It also
offers new insights on the necessary conditions for their validity: a physical contribution to the vertex functional must satisfy certain analyticity properties
in the soft limit in order for the consistency relations to hold. For standard inflationary models, this is equivalent to requiring that mode functions have constant growing-mode solutions.
For more exotic models in which modes do not ``freeze'' in the usual sense, the analyticity requirement offers an unambiguous criterion.
\end{abstract}

\newpage
\setcounter{page}{1}

\renewcommand{\thefootnote}{\arabic{footnote}}
\setcounter{footnote}{0}

\linespread{1.1}
\parskip 4pt

\section{Introduction}

In recent years there has been interest in consistency relations for primordial perturbations. The simplest one~\cite{Maldacena:2002vr,Creminelli:2004yq,Cheung:2007sv,Senatore:2012wy,Creminelli:2012ed,Hinterbichler:2012nm,Assassi:2012zq,Goldberger:2013rsa,Hinterbichler:2013dpa} relates an $N$-point correlation function with a soft scalar operator $\zeta$ to a scale transformation of the $N-1$-point correlation function without the soft mode\footnote{Here $\zeta$ is the curvature perturbation in uniform-density gauge~\cite{Bardeen:1983qw,Salopek:1990jq}, $P_\zeta$ is the power spectrum, $\la \ldots \ra'$ are correlators without the momentum-conserving $\delta$-function, and the subscript $c$ denotes the connected part. See the main text for further details.}:
\be 
\lim_{\vec{q}\rightarrow 0} \frac{1}{P_\zeta(q)} \la  \zeta(\vec{q}) {\cal O}(\vec{p}_1,\ldots,\vec{p}_N)\ra_c'   =  
- \Bigg( 3(N-1)  +   \sum_{a=1}^N  \vec{p}_a\cdot \frac{\partial}{\partial \vec{p}_a}\Bigg)\la   {\cal O}(\vec{p}_1, \ldots,\vec{p}_N)\ra_c'   \,.
\label{warddil}
\ee
Here, ${\cal O}(\vec{p}_1,\ldots,\vec{p}_N)$ is an arbitrary equal-time product of scalar $\zeta$ and tensor $\gamma_{ij}$ modes, with momenta $\vec p_1, \ldots, \vec p_N$. There is an analogous relation involving a soft tensor $\gamma_{ij}(\vec{q})$ related to an anisotropic rescaling of the lower-point function~\cite{Maldacena:2002vr}. The power of these relations lie in their generality: {\it Any} early universe scenario involving a single scalar degree of freedom (or single `clock'), and whose perturbations become constant at late times, must satisfy~(\ref{warddil}). Conversely, they can be violated if multiple fields contribute to density perturbations and/or $\zeta$ grows outside the horizon~\cite{Cai:2009fn,Khoury:2008wj,Namjoo:2012aa,Chen:2013aj,Chen:2013eea}.\footnote{The consistency relation~(\ref{warddil}) can be interpreted as the statement that primordial correlation functions in a suitable coordinate system vanish in the soft limit~\cite{Tanaka:2011aj,Pajer:2013ana}. Nevertheless, this statement is not tautological ---  the fact that the consistency relations can be violated implies that they constitute physical, observationally-testable probes of early universe physics.}

The consistency relation~(\ref{warddil}) is a consequence of symmetry: it follows from the Ward identity for spontaneously broken spatial dilations~\cite{Maldacena:2002vr,Assassi:2012zq,Goldberger:2013rsa,Hinterbichler:2013dpa}.\footnote{See~\cite{Schalm:2012pi,Bzowski:2012ih} for recent derivations of~(\ref{warddil}) using holographic arguments.} More generally, scalar perturbations on {\it any} spatially-flat cosmological background non-linearly realize the full conformal group $SO(4,1)$ on $\mathds{R}^3$~\cite{Creminelli:2012ed,Hinterbichler:2012nm}. The origin of conformal symmetry is most easily seen in comoving gauge, where the spatial metric (ignoring tensors) $h_{ij} = a^2(t) e^{2\zeta(\vec{x},t)}\delta_{ij}$ is conformally flat and hence invariant under the 10 conformal transformations on $\mathds{R}^3$. The symmetry breaking pattern is $SO(4,1)\rightarrow {\rm spatial} \; {\rm rotations} + {\rm translations}$, with $\zeta$ playing the role of the Goldstone boson (or dilation) for the broken dilation and special conformal transformations (SCTs). The Ward identities associated with the SCTs also give rise to consistency relations~\cite{Creminelli:2012ed}. These relate the order $q$ behavior of an $N+1$-point correlation function with a soft $\zeta$ mode to a SCT on the $N$-point function. 

Recently, it has been shown that cosmological perturbations enjoy an infinite number of non-linearly realized global symmetries~\cite{Hinterbichler:2013dpa}. 
These are residual diffeomorphisms mapping field configurations which fall off at infinity into those which do not. Nevertheless, certain linear combinations
of these transformations can be smoothly extended to physical configurations which fall off at infinity, and as such constitute adiabatic modes~\cite{Weinberg:2003sw}.
These symmetries can be labeled by an integer $n\geq 0$ and involve both scalar and tensor perturbations. 

The corresponding Ward identities imply {\it an infinite number of consistency relations}~\cite{Hinterbichler:2013dpa}, of which~(\ref{warddil}) is the simplest case. At each order, they constrain --- completely for $n=0,1$, and partially for $n\geq 2$ --- the $q^n$ behavior of an $N+1$-point correlation function with a soft scalar or tensor mode to a symmetry transformation on an $N$-point function. Schematically, they are of the form
\be
\lim_{\vec{q}\rightarrow 0} {\partial^n \over \partial q^n}
\left( {1\over P_\zeta (q)} \langle \zeta(\vec q) {\cal O}\rangle'_c + 
{1\over P_\gamma (q)} \langle \gamma(\vec q) {\cal O}\rangle'_c\right)
\sim - {\partial^n \over \partial p^n}  \langle {\cal O}
  \rangle'_c \, .
\label{schematic}
\ee
There are 3 independent relations for $n=0$ (the dilation identity~(\ref{warddil}) involving a soft scalar, and two involving
a soft tensor), 7 relations for $n=1$ (the 3 SCT identities involving a soft scalar, and 4 involving a soft tensor), and 6 for each $n\geq 2$ (4 involving a soft tensor, and 2 involving
mixtures of soft scalar and tensor). The $n=0$ and $n=1$ relations were known from background-wave arguments. The $n\geq 2$ relations were discovered in~\cite{Hinterbichler:2013dpa}.

In this paper, we show that the consistency relations~(\ref{schematic}) all derive from a single, {\it master identity}, which follows from the Slavnov-Taylor identity for spatial diffeomorphisms.
This master identity is valid at any value of $q$ and therefore goes beyond the soft limit. By differentiating it $n$ times with respect to $q$ and setting $q= 0$, we recover~(\ref{schematic})
at each order. Our approach underscores the role of diffeomorphism invariance at the root of cosmological consistency relations. It also offers insights on the necessary
assumptions for their validity.

We will derive the master identity in two independent ways: first, using the fixed-time path-integral approach introduced in~\cite{Goldberger:2013rsa}; second, using the 4d path integral.
For simplicity, we focus here on soft 3-point functions, with the hard momenta given by scalar modes. The generalization to more general correlation functions should be straightforward. 
Let us illustrate the results in the fixed-time approach, for concreteness. The master identity takes the form
\beq
\frac{1}{3}q_i\Gamma^{\zeta\zeta\zeta}(\vec{q},\vec{p},-\vec{q}-\vec{p})+2 q^j\Gamma_{ij}^{\gamma\zeta\zeta}(\vec{q},\vec{p},-\vec{q}-\vec{p})=q_i \Gamma_\zeta( p )-p_i\bigg( \Gamma_\zeta(|\vec{q}+\vec{p}|)-\Gamma_\zeta( p ) \bigg)\,,
\label{wardfintro}
\eeq
where $\Gamma^{\zeta\zeta\zeta}$ and $\Gamma_{ij}^{\gamma\zeta\zeta}$ are respectively the cubic vertex functions for 3 scalars, and for 2 scalars$-$1 tensor, each without the momentum-conserving delta function, while $\Gamma_\zeta$ is the inverse scalar propagator.\footnote{Note that $\Gamma_{ij}^{\gamma\zeta\zeta}$ is traceless ($\delta^{ij}\Gamma_{ij}^{\gamma\zeta\zeta} = 0$), but not necessarily transverse ($q^j\Gamma_{ij}^{\gamma\zeta\zeta}\neq 0$).} The solution for $\frac{1}{3}\delta_{ij}\Gamma^{\zeta\zeta\zeta}(\vec{q},\vec{p},-\vec{q}-\vec{p})+2 \Gamma_{ij}^{\gamma\zeta\zeta}(\vec{q},\vec{p},-\vec{q}-\vec{p})$ can be obtained as a power series around $q = 0$, up to an arbitrary symmetric, transverse matrix $A_{ij}$. This arbitrary term is model-dependent, and hence contains physical information about the underlying theory. It stems from the fact that~(\ref{wardfintro}) only constrains the longitudinal components of the vertex functions. {\it The key assumption underlying the consistency relations is that $A_{ij}$ is analytic in $q$, specifically that it starts at ${\cal O}(q^2)$}. For standard inflationary scenarios, we will see that this is equivalent to the usual assumption of constant asymptotic solutions for the mode functions. For more exotic examples, such as khronon inflation~\cite{Creminelli:2012xb}, our criterion is the unambiguous one.

Up to $q^2$ order, therefore, we can isolate $\Gamma^{\zeta\zeta\zeta}$,
and then convert to correlation functions as usual: $\langle \zeta_{\vec{p}_1}\zeta_{\vec{p}_2} \zeta_{\vec{p}_3}\rangle = P_\zeta(p_1) P_\zeta( p_2 ) P_\zeta (p_3)\Gamma^{\zeta\zeta\zeta}(\vec{p}_1,\vec{p}_2,\vec{p}_3)$.
The result is
\beq
\frac{\langle \zeta_{\vec{q}}\zeta_{\vec{p}} \zeta_{-\vec{p}-\vec{q}} \rangle'}{P_\zeta(q)}&=&-\left(3+\vec{p}\cdot \frac{\p}{\p \vec{p}}\right)P_\zeta( p )\nonumber \\
&-&\frac{1}{2}q^i\left( 6\frac{\p}{\p p^i} - p_i \frac{\p^2}{\p p^j \p p_j} +2p_j \frac{\p^2}{\p p^j \p p^i} \right)P_\zeta( p )+\mathcal{O}(q^2)\,.
\eeq
The first line is the dilation consistency relation and agrees with~(\ref{warddil}) for the case of interest. The second line reproduces the SCT consistency relation~\cite{Creminelli:2012ed}.
At order $q^2$ and higher, the physical term $A_{ij}$ contributes. However, its contribution can be removed order by order by taking linear combinations of $\langle \zeta\zeta\zeta\rangle$ and $\langle \gamma\zeta\zeta\rangle$, and taking a suitable projection. In this way, we will recover the general consistency relations~(\ref{schematic}). 

Interestingly, the essence of our method is fully captured by Quantum Electrodynamics (QED), which we present along the lines of~\cite{Slavnov} in Sec.~\ref{QEDeg}. The idea is simple: although the gauge is usually fixed by a gauge-fixing term, the gauge symmetry does give us an information about the interaction vertices. In other words, the gauge invariant part of the interaction vertices is constrained by the symmetry and the corrections coming from the gauge-fixing term can be accounted for explicitly. The resulting Ward-Takahashi identity~\cite{Ward:1950xp,Takahashi:1957xn}, as is well known, relates the (longitudinal part of the) photon-fermion vertex to the fermion propagator. By expanding this identity as a power series in the soft photon momentum $q$, we will obtain QED consistency relations analogous to ones obtain in \cite{Hinterbichler:2013dpa} for cosmology.

The approach described in this paper is quite general and can be applied beyond the early universe, for instance, to derive consistency relations for the large scale structure~\cite{Kehagias:2013yd,Peloso:2013zw,Creminelli:2013mca,Lamupcoming}. It can also be applied to derive consistency relations for modified initial states~\cite{Holman:2007na,Meerburg:2009ys,Meerburg:2009fi,Ganc:2011dy,Chialva:2011hc,Agarwal:2012mq,Flauger:2013hra,Ashoorioon:2013eia,Aravind:2013lra}, as well as consistency relations for multiple soft limits~\cite{Austinupcoming}. Another interesting arena is the conformal mechanism~\cite{Rubakov:2009np,Creminelli:2010ba,Hinterbichler:2011qk,Hinterbichler:2012mv,Creminelli:2012my,Hinterbichler:2012fr,Hinterbichler:2012yn}, whose consistency relations have been derived recently~\cite{Creminelli:2012qr}. 

The paper is organized as follows. In Sec.~\ref{QEDeg}, we begin with the warm-up example of QED and outline all of its relevant properties. In Sec.~\ref{fixedt}, we turn to the derivation of the Slavnov-Taylor identity for cosmological perturbations, first using the fixed-time/three-dimensional Euclidian path-integral method proposed in~\cite{Goldberger:2013rsa}. In Sec.~\ref{twohard}, we illustrate how the consistency relations derive from the Slavnov-Taylor identity, focusing for simplicity on consistency relations involving two hard scalar modes with a soft scalar or tensor mode. In Sec.~\ref{full4d}, we rederive these results, this time using the conventional four-dimensional in-in path integral. Some of technical details of our derivations have been relegated to a series of Appendices. We summarize the results in Sec.~\ref{conclu} discuss further applications of the method outlined here.

\section{Ward Identities for Electrodynamics}
\label{QEDeg}

As an abelian warm-up to the cosmological case, consider the Ward-Takahashi identities for QED~\cite{Ward:1950xp,Takahashi:1957xn}, derived as a consequence of gauge symmetry~\cite{Slavnov} instead of its global subgroup. 
The generating functional for QED with a single Dirac fermion $\psi$ is given by the following path integral
\beq
Z[J_\mu,\eta,\bar{\eta}]=\int \mathcal{D}A_\mu \mathcal{D}\bar{\psi}\mathcal{D}\psi e^{ i S + iS_{\rm g.-f.} + iS_{\rm ext.}}\,,
\label{QEDGen}
\eeq
where $S[A_\mu,\bar{\psi},\psi]$ is the gauge invariant QED action, and 
\be
S_{\rm g.-f.}[A_\mu]= -\frac{1}{2\xi}  \int {\rm d}^4x (\p^\mu A_\mu)^2 \,;\qquad  S_{\rm ext.}[A_\mu,\bar{\psi},\psi] = \int {\rm d}^4x  \left(J^\mu A_\mu+ \bar{\eta}\psi+\bar{\psi}\eta\right)
\ee
are the gauge-fixing term and external current contributions, respectively.\footnote{Here, $\xi$ is the standard gauge parameter, {\it e.g.} $\xi = 1$ is Feynman gauge.}

To derive the Ward-Takahashi identities, we perform an infinitesimal gauge transformation\footnote{The fermion is assumed to carry a unit charge.}
\beq
A_\mu\rightarrow A_\mu+\p_\mu \Lambda\,; \qquad \psi\rightarrow\psi-i\Lambda \psi\,,
\label{gaugetran}
\eeq
where $\Lambda(x)$ is an infinitesimal gauge parameter. Since $S$ and the integration measure are both gauge invariant,
the variation of the generating functional is 
\beq
\nonumber
\delta Z[J_\mu,\eta,\bar{\eta}] & =& i \int \mathcal{D}A_\mu \mathcal{D}\bar{\psi}\mathcal{D}\psi e^{ i S + iS_{\rm g.-f.} + iS_{\rm ext.}}  \int {\rm d}^4x\, \Lambda(x)\left[ -\frac{\Box}{\xi}\p^\mu A_\mu-\p^\mu J_\mu-i\bar{\eta}\psi + i\bar{\psi}\eta \right] \\
&=&i \int {\rm d}^4x\, \Lambda(x) \left[ \frac{i\Box}{\xi} \p^\mu \frac{\delta}{\delta J^\mu}-\p^\mu J_\mu-\bar{\eta}\frac{\delta}{\delta\bar{\eta}}+\eta\frac{\delta}{\delta\eta} \right]Z[J,\bar{\eta},\eta] \,.
\label{delZQED}
\eeq
The generating functional should be invariant ($\delta Z = 0$) under~(\ref{gaugetran}), since it is merely a field redefinition. Since $\Lambda(x)$ is arbitrary, this leads to the functional differential equation
\beq
\left[ \frac{i\Box}{\xi} \p^\mu \frac{\delta}{\delta J^\mu}-\p^\mu J_\mu-\bar{\eta}\frac{\delta}{\delta\bar{\eta}}+\eta\frac{\delta}{\delta\eta} \right]Z[J,\bar{\eta},\eta]=0\,.
\label{WardZ}
\eeq
Clearly, the generating functional of connected diagrams, $W \equiv - i \ln Z$, obeys a similar differential equation. Performing the standard Legendre transform to the vertex functional $\Gamma\equiv W - S_{\rm ext.}$, and
using the standard relations $J^\mu = - \frac{\delta \Gamma}{\delta A_\mu}$, $A_\mu = \frac{\delta W}{\delta J^\mu}$, {\it etc.}, this implies
\beq
-\frac{\Box}{\xi}\p^\mu A_\mu+\p_\mu\frac{\delta \Gamma}{\delta A_\mu}+i\psi\frac{\delta\Gamma}{\delta\psi}-i\bar{\psi}\frac{\delta\Gamma}{\delta\bar{\psi}}=0\,.
\label{qedgamma}
\eeq
Note that nowhere did we need the explicit form of the QED action $S$ --- all we used was its invariance under~(\ref{gaugetran}). Therefore, the identity~(\ref{qedgamma}) holds more generally for {\it any} gauge invariant action.
We will henceforth assume this most general situation.

By varying~(\ref{qedgamma}) a number of times with respect to the fields, and setting the fields to zero after variation, one can obtain various relations among the vertices of the theory. 
For instance, varying with respect to $\psi$ and $\bar{\psi}$ gives (in momentum space):
\beq
q^\mu \Gamma_\mu^{A\bar{\psi}\psi}(q,p,-p-q)=\Gamma_\psi(p+q)-\Gamma_\psi( p )\,,
\label{QEDWard}
\eeq
where $\Gamma_\mu^{A\bar{\psi}\psi} =  \frac{\delta^3 \Gamma}{\delta A^\mu \delta\psi\delta\bar{\psi}}$ is the three-point vertex, and $\Gamma_\psi ( p )  =   \frac{\delta^2 \Gamma}{\delta\psi\delta\bar{\psi}}$
is the inverse fermion propagator. Equation~(\ref{QEDWard}) is the celebrated Ward-Takahashi identity~\cite{Ward:1950xp,Takahashi:1957xn}. It exhibits the constraint that must be obeyed by the vertex functionals. 

We are interested in deriving the identity for correlation functions, which are related to the vertex functionals as follows:
\beq
\nonumber
&&P_\psi(p)\equiv \langle \bar{\psi}_p \psi_{-p} \rangle'=1/\Gamma_\psi(p), \\
&&\langle A^\mu_q \bar{\psi}_p \psi_{-p-q} \rangle' =  -\langle A^\mu_q A^\nu_{-q} \rangle' P_\psi (p) P_{{\psi}}(p+q)\Gamma_\nu^{A\bar{\psi}\psi}(q,p,-p-q)\, ,
\eeq
where primes indicate correlators with the delta function removed:
\be
\langle {\cal O}(k_1,\ldots,k_N) \rangle \equiv (2\pi)^4\delta^4(k_1 + \ldots + k_N) \langle {\cal O}(k_1,\ldots,k_N) \rangle'\,.
\ee
This makes clear that the quantity we would like to solve for is $P_\psi (p) P_{{\psi}}(p+q)\Gamma_\nu^{A\bar{\psi}\psi}(q,p,-p-q)$. 
Rewriting~(\ref{QEDWard}) in terms of this quantity, we obtain
\beq
q^\mu P_\psi (p) P_{{\psi}}(p+q)\Gamma_\mu^{A\bar{\psi}\psi}(q,p,-p-q)=P_\psi(p)-P_\psi( p+q )\,.
\eeq

It is straightforward to see that the most general solution to this equation is given by 
\beq
P_\psi (p) P_{{\psi}}(p+q)\Gamma_\mu^{A\bar{\psi}\psi}(q,p,-p-q)=-\frac{\p P_\psi( p )}{\p p^\mu} - \sum_{n=1}^{\infty} \frac{q^{\alpha_1}\ldots q^{\alpha_n}}{(n+1)!} \frac{\p^{n+1}P_\psi( p )}{\p p^\mu \p p^{\alpha_1}\ldots \p p^{\alpha_n}}+C_\mu\,,
\label{QEDexp}
\eeq
where $C_\mu$ is an arbitrary transverse vector,
\be
q^\mu C_\mu=0\,. 
\label{Ctran}
\ee
Its more general form is therefore
\be
C_\mu = (q^2 \eta_{\mu\alpha}-q_\mu q_\alpha)v^\alpha(q,p) + q^\alpha M_{\mu\alpha}(q,p)\,,
\label{Cdecomp}
\ee
where $v^\mu$ is an arbitrary vector, and $M_{\mu\nu}$ is anti-symmetric. The vector $C_\mu$ represents the part of the cubic vertex which is not fixed by symmetry arguments only, but depends on the details of the theory. It therefore encodes physical information
about the theory. 

Now comes the key assumption: if the theory is {\it local}, which is of course the case for QED, then $C_\mu$ {\it should be analytic in $q$.} In terms of the general decomposition~(\ref{Cdecomp}),
this implies that both $v^\mu$ and $M_{\nu\mu}$ start at order $q^0$. In particular, it follows that $C_\mu$ does not contribute at $q^0$ order, thus the cubic vertex is determined
by the derivative of the inverse fermion propagator at leading order:
\beq
\Gamma_\mu^{A\bar{\psi}\psi}(0,p,-p)=\frac{\p \Gamma_\psi( p )}{\p p^\mu}\,.
\label{QEDWardq0}
\eeq
This is the QED analogue of Maldacena's consistency relation~(\ref{warddil}).\footnote{Notice that~\eqref{QEDWardq0} is satisfied not only for fermion QED, but for scalar QED as well. This traces back to the fact that fermions and scalars have identical transformation properties under the gauge symmetry.}

At the next order in $q$, however, $C_\mu$ can contribute through the $M_{\mu\nu}$ term. For instance, with fermions it can take the form 
$C^\mu= q_\nu [\gamma^\nu,\gamma^\mu]$, where the $\gamma$'s are the usual Dirac matrices. This will arise if the theory includes an anomalous magnetic dipole interaction, $F_{\mu\nu}\bar{\psi}\gamma^\mu\gamma^\nu\psi$.
More generally, we see that $C_\mu$ encodes information about non-minimal photon couplings to the fermions. It vanishes identically for QED, where the photon-fermion coupling is minimal. 

We can translate the identity~(\ref{QEDexp}) to a statement about correlation functions by contracting the vertex functional with the appropriate Green's functions. Specializing to Lorentz gauge, $\partial^\mu A_\mu = 0$, we have 
\beq
\frac{\langle A^\mu_q \bar{\psi}_p \psi_{-p-q} \rangle'}{P_A(q)} = - P^{\mu\nu}(\hat{q}) P_\psi(p)P_\psi(p+q)\Gamma_\nu^{A\bar{\psi}\psi}(q,p,-p-q)\,,
\eeq
where $P_A$ is defined as $\langle A^\mu_q A^\nu_{-q}\rangle '=P_A(q) P^{\mu\nu}(\hat{q})$, with $P_{\mu\nu}(\hat{q})= \eta_{\mu\nu} - \hat{q}_\mu\hat{q}_\nu$, $\hat{q}^\mu \equiv q^\mu/q$, denoting the transverse projector.

Ideally we would like to derive model-independent ({\it i.e.}, $C_\mu$-independent) relations among the correlation functions. To do so, we must apply suitable component operators
$P_{\mu \ell_1\cdots \ell_n \nu m_1 \ldots m_n}$ on each term of the Taylor series~(\ref{QEDexp}) such that $C_\mu$ is projected out at each order. In other words, the $C_\mu$ contribution
on the right-hand side of
\ba
\nonumber
& & \lim_{q\rightarrow 0} P_{\mu \ell_1\cdots \ell_n \nu m_1 \ldots m_n}(\hat{q}) \frac{\partial^n}{\partial q_{m_1}\cdots \partial q_{m_n}} \bigg[\frac{\langle A^\nu_q \bar{\psi}_p \psi_{-p-q} \rangle'}{P_A(q)} \bigg]= \\
& &  - \lim_{q\rightarrow 0} P_{\mu \ell_1\cdots \ell_n \nu m_1 \ldots m_n}(\hat{q}) \frac{\partial^n}{\partial q_{m_1}\cdots \partial q_{m_n}}\bigg( P^{\nu\alpha}(\hat{q}) P_\psi(p)P_\psi(p+q)\Gamma_\alpha^{A\bar{\psi}\psi}(q,p,-p-q)\bigg)
\label{qedlim}
\ea
should drop out. We claim this is achieved if $P_{\mu \ell_1\cdots \ell_n \nu m_1 \ldots m_n}$ is:

\begin{enumerate}

\item Symmetric in the $(\mu,\ell_1,\ldots,\ell_n)$ indices and in the $(\nu,m_1 \ldots m_n)$ indices.

\item Symmetric under the interchange of sets of indices: $(\mu,\ell_1,\ldots,\ell_n) \leftrightarrow (\nu,m_1 \ldots m_n)$.

\item Traceless:
\be
P^\mu_{~\mu \ell_2\cdots \ell_n \nu m_1 \ldots m_n}(\hat{q}) = 0\,.
\label{qedp1}
\ee

\item Transverse with respect to $q$:
\beq
q^\mu P_{\mu \ell_1\cdots \ell_n \nu m_1 \ldots m_n}(\hat{q}) = 0\,.
\label{qedp2}
\eeq

\end{enumerate}
In Appendix A, we show explicitly that $C_\mu$ is indeed successfully projected out by the projector defined above.\footnote{The properties of this projector can be further motivated through an ``adiabaticity'' argument similar to that given in~\cite{Hinterbichler:2013dpa} for the cosmological case. The Lorentz gauge condition $\partial^\mu A_\mu = 0$ is preserved by gauge transformations $\delta A_\mu = \partial_\mu \lambda$, where $\lambda$ is a harmonic function: $\Box \lambda = 0$. Expanding as a Taylor series about the origin, we have $\lambda(x) = \sum_{n=0}^\infty \frac{1}{(n+1)!}M_{\ell_0 \ldots \ell_n}  x^{\ell_0}\cdots x^{\ell_n}$, where the array $M_{\ell_0 \ldots \ell_n}$ is fully symmetric and traceless. (We have ignored the constant term in the expansion, since it leaves $A_\mu$ invariant.)
In momentum space, this generates 
\be
\delta A_\mu(q) = \frac{(-i)^n}{n!} M_{\mu \ell_1 \ldots \ell_n}  \frac{\partial^{n}}{\partial q_{\ell_1}\cdots \partial q_{\ell_n}}(2\pi)^4\delta^4(q)\,.
\ee
For this configuration to be extendible to a {\it physical} mode, with suitable fall-off behavior at spatial infinity, we imagine smoothing out the momentum profile around $q = 0$. To ensure
that transversality is preserved in Fourier space at finite momentum, $q^\mu \delta A_\mu = 0$, we must let the $M_{\mu \ell_1 \ldots \ell_n}$ coefficients become $\hat{q}$-dependent such that
\be
q^\mu M_{\mu \ell_1 \ldots \ell_n}(\hat{q})  = 0\,.
\ee
In other words, $M_{\mu \ell_1 \ldots \ell_n}$ is fully symmetric, traceless and transverse. The corresponding projector $P_{\mu \ell_1\cdots \ell_n \nu m_1 \ldots m_n}(\hat{q})$ appearing in the identities must therefore satisfy the properties listed in the main text.}

Making use of this fact when substituting~(\ref{QEDexp}) into~(\ref{qedlim}), we obtain the following consistency relations
\beq
& & \lim_{q\rightarrow 0} P_{\mu \ell_1\cdots \ell_n \nu m_1 \ldots m_n}(\hat{q}) \frac{\partial^n}{\partial q_{m_1}\cdots \partial q_{m_n}} \bigg[\frac{\langle A^\nu_q \bar{\psi}_p \psi_{-p-q} \rangle'}{P_A(q)} \bigg]=-\frac{P_{\mu \ell_1\cdots \ell_n \nu m_1 \ldots m_n}(\hat{q})}{n+1}\frac{\p^{n+1} P_\psi}{\p p_\nu\p p_{m_1}\ldots \p p _{m_n}}.\nonumber
\eeq
These are very similar in form to the Ward identities for cosmological perturbations derived in~\cite{Hinterbichler:2013dpa}. In the following sections we will reproduce these identities as a consequence of spatial diffeomorphism invariance. The derivation is closely analogous to the one above, with the replacement of $U(1)$ gauge symmetry by diffeomorphism invariance. Because of the non-Abelian nature of the latter we will refer to the resulting identities similar as Slavnov-Taylor identities for cosmology.

\section{Slavnov-Taylor Identities for Cosmology}
\label{fixedt}

We now turn to the derivation of cosmological consistency relations. Our method follows Slavnov's classic work~\cite{Slavnov}, applied to cosmology.\footnote{For the flat space considerations, see~\cite{Medrano}.} The non-abelian nature of the symmetries of interest (namely, the diffeomorphism invariance of GR) complicates the derivation to some extent. In particular, the gauge-fixing term, which in the abelian case dropped out of the identity~\eqref{QEDWard}, does contribute to the Slavnov-Taylor identities in the non-abelian case. As shown in Appendix B, however, the gauge-fixing term only contributes at loop order. While this contribution can be accounted for explicitly if desired, we avoid the unnecessary complications and work at tree-level. 

For simplicity, we begin the demonstration of our method in the framework of the fixed-time path-integral formalism of~\cite{Goldberger:2013rsa}.
The basic idea is simple: since we are solely interested in correlation functions of fields evaluated at the final time (as opposed to unequal-time correlators,
or correlators involving time-derivatives of the fields), it is convenient to work with a three-dimensional Euclidean path integral over field
configurations at the final time, with the ``history'' information being encoded in the wavefunction. The fixed-time formalism makes the derivation
simpler and more transparent. In Sec.~\ref{full4d}, we will reproduce the same results using the four-dimensional in-in path integral.

We consider the diffeomorphism invariant theory of the metric degrees of freedom $g\mn$ and inflaton $\phi$, around the
spatially-flat Friedmann-Robertson-Walker background
\beq
\bar{g}\mn {\rm d}x^\mu {\rm d}x^\nu=-{\rm d}t^2+a^2(t){\rm d}\vec{x}^2\,; \qquad  \phi=\bar{\phi}(t)\,,
\eeq
and parameterize the excitations as
\beq
g\mn=\bar{g}\mn (t)+a^2(t)h\mn\,; \qquad  \phi=\bar{\phi}(t)+\f\,.
\eeq
According to~\cite{Goldberger:2013rsa}, the correlation functions at fixed time $t$ can be conveniently described by the Euclidean generating functional
\beq
\nonumber
Z[T,J]&=&\int \mathcal{D}h_{ij} \mathcal{D}\f \left\vert \Psi[h,\f, t]\right\vert^2 e^{S_{\rm ext.}}\,; \\
S_{\rm ext.} &=& \int {\rm d}^3 x \left(h_{ij} T^{ij}+\f J\right)\,,
\label{3d}
\eeq
where $T_{ij}$ and $J$ represent tensor and scalar currents, respectively, while $\Psi[h,\f,t]$ is the wavefunctional at time $t$.\footnote{The peculiarities related to the gauge-fixing term will be addressed separately and can be ignored for the moment.} Note that the auxiliary fields $h_{00}$ and $h_{0i}$ (equivalently, the lapse function and shift vector) have been integrated out using the constraint equations~\cite{Maldacena:2002vr}, so that the path integral is over the spatial metric $h_{ij}$ only.  

Since time is fixed in this approach, the time re-parametrization symmetry is explicitly broken by the formalism. The symmetries at hand are spatial diffeomorphisms $x^i\rightarrow x^i -\xi^i$, under which the fields transform as
\beq
h_{ij} & \rightarrow & h_{ij}+\p_i \xi_{j}+\p_j \xi_{i}+\xi^k \p_k h_{ij}+h_{ik}\p_j\xi^{k}+h_{jk}\p_i \xi^{k}\,;\nonumber \\
\f & \rightarrow & \f+\p_k \f~\xi^k\,.
\label{sym}
\eeq
From now on, all the indices are assumed to be raised and lowered using $\delta_{ij}$. Analogously to the QED case --- see~(\ref{delZQED}) ---, the invariance of the generating functional under this field redefinition leads to
\beq
\nonumber
0 & = & \int \mathcal{D}h_{ij} \mathcal{D}\f \left\vert \Psi[h,\f, t]\right\vert^2 e^{S_{\rm ext.}} \int {\rm d}^3x~ \xi^k\bigg\{\text{(G.F.)}_k -2 \p_j T^{j}_{k}+\p_k h_{ij} T^{ij}-2\p_j \left( h_{ik} T^{ij} \right)+\p_k \f J \bigg\} \\
\nonumber
&=&  \int {\rm d}^3x~ \xi^k \left\{ \text{(G.F.)}_k-2 \p_j T^{j}_k+\p_k \left(\frac{\delta}{\delta T^{ij}}\right) T^{ij}-2\p_j \left(\frac{\delta}{\delta T^{ik}} T^{ij} \right)+\p_k \left(\frac{\delta}{\delta J}\right) J \right\}Z[T,J] \,,\\
\label{sti}
\eeq
where in the last step we have made the replacements $\f\rightarrow \frac{\delta}{\delta J}$ and $h_{ij}\rightarrow \frac{\delta}{\delta T^{ij}}$. Here, $\text{(G.F.)}_k$ denotes terms arising from the variation of the gauge-fixing term; we will be schematic about it until its explicit form becomes important.
Since $\xi^k$ is arbitrary, the integrand itself must vanish. Rewriting the result in terms of $W = \ln Z$, the generator of connected amplitudes, we obtain 
\beq
\text{(G.F.)}_k -2 \p_j T^{j}_k(x)+\p_k \left(\frac{\delta W}{\delta T^{ij}(x)}\right) T^{ij}(x)-2\p_j \left( \frac{\delta W}{\delta T^{ik}(x)} T^{ij}(x) \right)+\p_k \left(\frac{\delta W}{\delta J(x)}\right) J(x)=0 \,.
\label{ward1} 
\eeq

We can convert~(\ref{ward1}) to an equation for the vertex functional by means of the Legendre transform
\beq
\Gamma[h,\f]=W[T,J]-\int {\rm d}^3 x ~(h_{ij} T^{ij}+\f J)\,,
\eeq
which implies
\beq
\frac{\delta\Gamma}{\delta h_{ij}}&=&-T^{ij}\,; \qquad \frac{\delta W}{\delta T^{ij}}=h_{ij}\,;\\
\frac{\delta\Gamma}{\delta \f}&=&-J\,; \qquad~~~ \frac{\delta W}{\delta J}=\f\,.
\eeq
The resulting equation for $\Gamma$ is
\beq
\text{(G.F.)}_k+2 \p_j \frac{\delta \Gamma}{\delta h_{jk}}=(\p_k h_{ij}) \frac{\delta\Gamma}{\delta h_{ij}}-2\p_j \left( h_{ik} \frac{\delta\Gamma}{\delta h_{ij}} \right)+\p_k \f \frac{\delta\Gamma}{\delta \f}.
\label{vertexward}
\eeq

At this point, we specialize to comoving gauge (or `$\zeta$ gauge'), defined by 
\beq
\varphi = 0\,;\qquad H_{ij}\equiv \delta_{ij}+h_{ij}=e^{2\zeta}\hh_{ij} \,,~~~\text{with}~~ \det\hh=1\,.
\eeq
It follows that
\beq
\nonumber
\zeta&=&\frac{1}{6}\ln\det H\,;\\
\hh_{ij}&=&\frac{H_{ij}}{(\det H)^{1/3}}\,.
\label{decomb}
\eeq
The variational derivative can be converted to the new variables using
\beq
\nonumber
\frac{\delta \Gamma}{\delta h_{ij}}&=& \frac{\delta \Gamma}{\delta\zeta}\frac{\delta\zeta(h)}{\delta h_{ij}}+\frac{\delta \Gamma}{\delta\hh_{k\ell}} \frac{\delta\hh_{k\ell}(h)}{\delta h_{ij}} \\
\nonumber
&=& e^{-2\zeta}\left\{\frac{1}{6}[\hh^{-1}]_{ij}\frac{\delta \Gamma}{\delta\zeta}+\left[ \delta_{ik}\delta_{j\ell}-\frac{1}{3}\hh_{k \ell}[\hh^{-1}]_{ij} \right]\frac{\delta \Gamma}{\delta\hh_{k\ell}}\right\} \\
&=& e^{-2 \zeta}\left\{ \frac{1}{6}\delta_{ij}\frac{\delta \Gamma}{\delta \zeta}+\frac{\delta \Gamma}{\delta \gamma_{ij}} \right\} + \ldots
\label{simpledecomb}
\eeq
where in the last step we have introduced
\be
\gamma_{ij} \equiv \ln \hh_{ij} \,;\qquad \gamma^i_{~i} = 0\,.
\ee
Here and henceforth, the ellipses indicate terms that are higher-order in $\gamma$. As we will see shortly, these will not contribute at tree-level to the consistency relations of interest in this paper. 

Substituting these results into~(\ref{vertexward}), the Slavnov-Taylor identity reduces to
\beq
\boxed{\frac{1}{\alpha} \Big(\vec{\nabla}^2 \p_j\gamma_{ij}+\p_i \p_j\p_k\gamma_{jk}\Big)+2\p_j \left( \frac{1}{6} \delta_{ij}\frac{\delta\Gamma}{\delta\zeta}+\frac{\delta \Gamma}{\delta \gamma_{ij}} \right)=\p_i \zeta \frac{\delta \Gamma}{\delta \zeta} + \ldots}
\label{ward}
\eeq
Note that we have only included the tree-level contribution of the gauge-fixing term explicitly, with $\alpha$ denoting a gauge-fixing parameter.
Hence this identity (and the consistency relations that derive from it) only holds at tree-level. This is one of the key results of this paper. By varying this identity a number of times with respect to $\zeta$ and $\gamma$, it is straightforward to obtain various consistency relations among the vertices of the theory. 

\section{Consistency Relations with Two Hard Scalar Modes}
\label{twohard}

We illustrate how consistency relations derive from~\eqref{ward} in the simplest case of a soft-momentum $\zeta$ or $\gamma$ mode coupled to two hard-momenta $\zeta$ modes. 
The generalization to higher-point correlation functions is straightforward. 

The consistency relations with two hard-momenta scalar modes are obtained by varying~\eqref{ward} with respect to $\zeta(\vec{x}_1)$ and $\zeta(\vec{x}_2)$ and then setting $\zeta=\gamma_{ij}=0$. Since the ellipses contain terms with powers of $\gamma$, they will all vanish upon setting $\gamma = 0$, as advocated. Expressing the result in Fourier space, we obtain\footnote{The relevant relations are
\beq
\nonumber
\int {\rm d}^3x_1\, {\rm d}^3x_2 e^{-i(\vec{p}_1\cdot\vec{x}_1 + \vec{p}_2\cdot \vec{x}_2)}\left.\frac{\delta^2 \Gamma}{\delta \zeta(\vec{x}_1)\delta \zeta(\vec{x}_2)}\right\vert_{\zeta = \gamma =0} &=& (2\pi)^3 \delta^3 (\vec{p}_1 + \vec{p}_2)\Gamma_\zeta( p_1 )\,;\\
\nonumber
\int {\rm d}^3 x_1 \, {\rm d}^3 x_2\, {\rm d}^3 x_3 e^{-i\sum \vec{p}_i\cdot \vec{x}_i}\left.\frac{\delta^3\Gamma}{\delta \zeta(\vec{x}_1) \delta \zeta(\vec{x}_2) \delta \zeta(\vec{x}_3)}\right\vert_{\zeta = \gamma =0} &=& (2\pi)^3\delta^3(\vec{p}_1+\vec{p}_2 + \vec{p}_3)\Gamma^{\zeta\zeta\zeta}(\vec{p}_1,\vec{p}_2,\vec{p}_3)\,;\\
\int {\rm d}^3 x_1 \, {\rm d}^3 x_2\, {\rm d}^3 x_3 e^{-i\sum \vec{p}_i\cdot \vec{x}_i} \left.\frac{\delta^3\Gamma}{\delta \gamma_{ij}(\vec{x}) \delta \zeta(\vec{y}) \delta \zeta(\vec{z})}\right\vert_{\zeta = \gamma =0}  &=& (2\pi)^3\delta^3(\vec{p}_1+\vec{p}_2 + \vec{p}_3)\Gamma_{ij}^{\gamma\zeta\zeta}(\vec{p}_1,\vec{p}_2,\vec{p}_3)\,.~~~~~
\eeq
}
\beq
\frac{1}{3}q_i\Gamma^{\zeta\zeta\zeta}(\vec{q},\vec{p},-\vec{q}-\vec{p})+2 q^j\Gamma_{ij}^{\gamma\zeta\zeta}(\vec{q},\vec{p},-\vec{q}-\vec{p})=q_i \Gamma_\zeta( p )-p_i\bigg( \Gamma_\zeta(|\vec{q}+\vec{p}|)-\Gamma_\zeta( p ) \bigg)\,.
\label{wardf}
\eeq
This is the master identity. By converting it to correlation functions, we will see how it recovers all known consistency relations involving a soft mode and two hard scalar modes.

We can translate~\eqref{wardf} to a statement about correlation functions using $\Gamma_\zeta ( p )=-P_\zeta^{-1}( p )$, as well as\footnote{Primed correlation functions are defined by removing the delta function:
\be
\langle {\cal O}(\vec{q}, \vec{k}_1, \ldots ,\vec{k}_N)\rangle  = (2\pi)^3\delta^3(\vec{q} +\vec{k}_1 + \ldots + \vec{k}_N) \langle {\cal O}(\vec{q}, \vec{k}_1, \ldots ,\vec{k}_N)\rangle' \,.
\label{primecor}
\ee
In particular, the power spectra are defined by
\beq
\nonumber
P_\zeta(k) &=& \langle \zeta_{\vec{k}}\zeta_{\vec{k}'} \rangle'\,; \\
P_\gamma(k) &=& \frac{1}{4} \delta_{ik}\delta_{j\ell} \langle \gamma^{ij}_{\vec{k}}\gamma^{k\ell }_{\vec{k}'}\rangle'\,. 
\eeq }
\beq
\nonumber
\langle \zeta_{\vec{q}} \zeta_{\vec{p}} \zeta_{-\vec{q} - \vec{p}} \rangle' &=&P_\zeta(q) P_\zeta( p )P_\zeta(|\vec{q} + \vec{p}|) \Gamma^{\zeta\zeta\zeta}(\vec{q},\vec{p},-\vec{q} - \vec{p}) \,;\\
\langle \gamma^{ij}_{\vec{q}}\zeta_{\vec{p}} \zeta_{-\vec{q} - \vec{p}} \rangle' &=& \hat{P}^{ijk\ell}(\hat{q})P_\gamma(q) P_\zeta( p )P_\zeta(|\vec{q} + \vec{p}|)\Gamma_{k\ell}^{\gamma\zeta\zeta}(\vec{q},\vec{p},-\vec{q} - \vec{p})\,,
\label{correl}
\eeq
where $\hat{P}_{ijk\ell}$ is the transverse, traceless tensor appearing in the graviton propagator:
\be
\hat{P}_{ijk\ell}(\hat{q}) = P_{ik} P_{j\ell} + P_{i\ell} P_{jk} - P_{ij}P_{k\ell} \,, 
\label{4proj}
\ee
with $P_{ij} \equiv \delta_{ij} - \hat{q}_i\hat{q}_j$ denoting the transverse projector. 

Proceeding analogously to the QED case, we can solve~\eqref{wardf}  as a Taylor series around $q = 0$:
\beq
P_\zeta(p )P_\zeta(|\vec{q}+\vec{p}|)\left(\frac{1}{3}\delta_{ij}\Gamma^{\zeta\zeta\zeta}(\vec{q},\vec{p},-\vec{q}-\vec{p})+2\Gamma_{ij}^{\gamma\zeta\zeta}(\vec{q},\vec{p},-\vec{q}-\vec{p})\right)= K_{ij} + A_{ij} \,,
\label{soln}
\eeq
where 
\beq
K_{ij} \equiv -\delta_{ij} P_\zeta( p ) -p_{(i}\frac{\p P_\zeta(p)}{\p p^{j)}}-\sum_{n=1}^{\infty} \frac{q_{\alpha_1}\ldots q_{\alpha_n}}{n!}\left[ \delta_{ij}\frac{\p ^n}{\p p_{\alpha_1}\ldots \p p_{\alpha_n}}+\frac{p_i}{n+1}\frac{\p ^{n+1}}{\p p_j \p p_{\alpha_1}\ldots \p p_{\alpha_n}}\right. \nonumber \\
\left.  +\frac{p_j}{n+1}\frac{\p ^{n+1}}{\p p_i \p p_{\alpha_1}\ldots \p p_{\alpha_n}} -\frac{p_{\alpha_1}}{n+1}\frac{\p ^{n+1}}{\p p_i \p p_j \p p_{\alpha_2}\ldots \p p_{\alpha_n}}\right]P_\zeta(p) \,,
\label{Kdef}
\eeq
and $A_{ij}$ is an arbitrary symmetric and transverse matrix:
\be
q^jA_{ij}=0\,.
\ee
Therefore it is of the general form\footnote{This can be seen by noting that the building blocks at our disposal are $\delta_{ij}$, $q_i$ and $p_i$. There is also the Levi-Cevita symbol $\epsilon_{ijk}$, but by symmetry $A_{ij}$ must be proportional to an even number of these, which is equivalent to products of $\delta$'s.}
\beq
A_{ij}=\epsilon _{ikm}\epsilon_{j\ell n}q^k q^\ell\bigg(a(\vec{p},\vec{q}) \delta^{mn}+b(\vec{p},\vec{q})p^mp^n\bigg)\,,
\label{ambig}
\eeq
where $a$ and $b$ are {\it a priori} arbitrary scalar functions of the momenta. The ambiguous nature of~\eqref{soln} originates from the form of~\eqref{wardf}, which only constrains the longitudinal components of the quantities at hand. Note that $A_{ij}$ is the analogue of $C_\mu$ for the QED case --- see~(\ref{QEDexp}) and~(\ref{Ctran}). This array encodes the physical part of the cubic vertices which are not fixed by symmetry arguments only, and hence depends on the details of the theory. 

Isolating the trace and traceless parts of~(\ref{soln}) allows us to solve for the individual vertices. 
Translating to correlation functions, we obtain
\beq
\nonumber
\frac{\langle \zeta_{\vec{q}} \zeta_{\vec{p}} \zeta_{-\vec{q} - \vec{p}} \rangle' }{P_\zeta(q)}&=& K+A \,;\\
\frac{\langle \gamma^{ij}_{\vec{q}}\zeta_{\vec{p}} \zeta_{-\vec{q} - \vec{p}} \rangle'}{P_\gamma(q)}&=&\frac{1}{2}\hat{P}^{ijk\ell }(\hat{q} )(K_{k\ell}+A_{k\ell}) \,.
\label{conv}
\eeq
To derive consistency relations from these, we must make an important assumption about the behavior of the arbitrary array $A_{ij}$ in the squeezed limit.

\subsection{Analyticity Assumption}
\label{analytic}

The key assumption for the validity of the consistency relations, as in the QED case, is that the functions $a$ and $b$ are analytic in $q$, {\it i.e.}, that
the physical term $A_{ij}$ starts at order $q^2$. This locality assumption on the effective action\footnote{The analyticity properties of $P_\zeta(p)P_{\zeta}(p+q)\Gamma$ and $\Gamma$ are obviously the same.} (or vertex functional $\Gamma$) is a non-trivial one: although GR is local by construction, recall that we are working in the framework where the lapse function and shift
vector have been integrated out, resulting in a spatially non-local action for $\zeta$ and $\gamma$.

In particular, let us see how this relates to the usual adiabaticity assumption, {\it i.e.}, that the growing mode solutions are constant.
Recall that non-local terms at cubic order arise from integrating out the shift vector, whose solution (at linear order) includes
\be
N_i  \supset -a^2\frac{\dot{H}}{H^2}  \frac{q_i}{q^2}\dot{\zeta}\,.
\ee
For the adiabatic mode, however, $\dot{\zeta}\propto q^2$, and this contribution becomes local.\footnote{The locality assumption was also implicit in~\cite{Goldberger:2013rsa},
for otherwise the effective action would be ill-defined at zero momentum. Indeed, in their approach one obtains
\beq
\int {\rm d}^3 k\, \delta^{(3)}(\vec{k})\frac{\delta\Gamma[\zeta]}{\delta\zeta_{\vec{k}}}=-\int{\rm d}^3 k \,\zeta_{\vec{k}} \, \vec{k}\cdot \frac{\partial}{\partial \vec{k}} \frac{\delta \Gamma[\zeta]}{\delta\zeta_{-\vec{k}} }\,,
\eeq
which can be used to derive  dilation consistency relation, provided of course that the integrals converge.} Conversely, in models where $\zeta$ is not constant outside the horizon (because of background instabilities~\cite{Khoury:2008wj}), the locality assumption is violated and the consistency relations will not hold. Similarly, this also explains why certain consistency relations fail in spatially non-local models, such as kronon inflation~\cite{Creminelli:2012xb}. 

In the remainder of the section we will see show how the consistency relations to all orders in $q$ follow from~(\ref{conv}), given the analyticity assumption.

\subsection{Recovering the order $q^0$ and $q$ consistency relations}

Since the physical term $A_{ij}$ kicks in at $q^2$ order, by assumption, the 3-point functions is uniquely determined by the 2-point function to zeroth and first order in $q$. 
The only contribution to the right-hand side of~\eqref{conv} at this order comes from $K_{ij}$, whose expansion is given by
\beq
K_{ij}=-\delta_{ij} P_\zeta( p )-p_{(i} \frac{\p P_\zeta}{\p p^{j)}}-\frac{1}{2}q_\ell \left[p_i\frac{\p^2 P_\zeta( p )}{\p p_j \p p_\ell}+p_j\frac{\p^2 P_\zeta( p )}{\p p_i \p p_\ell}-p_l\frac{\p^2 P_\zeta( p )}{\p p_i \p p_j}\right]+\mathcal{O}(q^2)\,.
\eeq
Substituting this into the first of~\eqref{conv} we obtain
\beq
\frac{\langle \zeta_{\vec{q}} \zeta_{\vec{p}} \zeta_{-\vec{q} - \vec{p}} \rangle' }{P_\zeta(q)}&=&-\left(3+p_k\frac{\p}{\p p_k}\right)P_\zeta( p )\nonumber \\
&-&\frac{1}{2}q_k\left( 6\frac{\p}{\p p_k}- p_k\frac{\p^2}{\p p_a \p p_a} +2p_a\frac{\p^2}{\p p_a \p p_k} \right)P_\zeta( p )+\mathcal{O}(q^2)\,.
\eeq
The first and second lines match respectively the dilation and SCT consistency relations~\cite{Creminelli:2012ed}.

Similarly, the second of~\eqref{conv} gives
\beq
\frac{\langle \gamma^{ij}_{\vec{q}}\zeta_{\vec{p}} \zeta_{-\vec{q} - \vec{p}} \rangle'}{P_\gamma(q)} &=& -\frac{1}{2}\hat{P}^{ijk\ell }(\hat{q}) p_k\frac{\p}{\p p_\ell} P_\zeta( p ) \nonumber \\
& + & \frac{1}{4}\hat{P}^{ijk\ell}(\hat{q}) q_m \left( p_m \frac{\p^2}{\p p_k \p p_\ell}-2p_k \frac{\p^2}{\p p_\ell \p p_m} \right)P_\zeta( p ) +   \mathcal{O}(q^2).
\eeq
The first and second lines correctly reproduce the anisotropic scaling~\cite{Maldacena:2002vr} and linear gradient tensor consistency relations~\cite{Creminelli:2012ed}, respectively.

\subsection{Higher-Order Consistency Relations}
\label{high}

At order $q^n$, $n\geq 2$, the soft correlation functions are only partially constrained by lower-point functions. The novel consistency relations with two hard scalar modes take the form~\cite{Hinterbichler:2013dpa}\footnote{The analogue of~(\ref{wardfinaln>2}) in~\cite{Hinterbichler:2013dpa} includes higher-order corrections in $\gamma_{ij}$. These do not appear here because we are working at tree-level.}
\beq
\nonumber
& &  \lim_{\vec{q}\rightarrow 0} P_{i\ell_0 \ldots \ell_n j m_0\ldots m_n}(\hat{q}) \frac{\partial^{n}}{\partial q_{m_1}\cdots \partial q_{m_n}} \Bigg(\frac{1}{P_\gamma(q)} \langle \gamma^{jm_0}_{\vec{q}}\zeta_{\vec{p}} \zeta_{-\vec{q} - \vec{p}}  \rangle' + \frac{\delta^{jm_0}}{3P_\zeta(q)} \langle  \zeta_{\vec{q}}\zeta_{\vec{p}} \zeta_{-\vec{q} - \vec{p}}  \rangle'  \Bigg) \\
&& =  - P_{i\ell_0 \ldots \ell_n j m_0\ldots m_n}(\hat{q}) \Bigg( \delta^{jm_0} \frac{\partial^{n}}{\partial p_{m_1} \cdots \partial p_{m_n}}  + \frac{p^{j}}{n+1}  \frac{\partial^{n+1}}{\partial p_{m_0} \cdots \partial p_{m_n}}\Bigg) P_\zeta( p ) \,.
\label{wardfinaln>2}
\eeq
The component operator $P_{i\ell_0\ldots \ell_n j m_0\ldots m_n}(\hat{q})$ has the following properties:

\begin{enumerate}

\item It is symmetric in the $(\ell_0,\ldots,\ell_n)$ indices and in the $(m_0 \ldots m_n)$ indices.

\item It is symmetric under the interchange of sets of indices: 
\be
P_{i\ell_0\ldots \ell_n  j m_0 \ldots m_n}(\hat{q}) = P_{j m_0 \ldots m_n i \ell_0\ldots \ell_n}(\hat{q})\,.
\ee

\item It obeys the trace condition:
\be
P_{i \ell\ell \ell_2\ldots \ell_n j m_0\ldots m_n}(\hat{q})  = - \frac{1}{3} P_{\ell i \ell \ell_2\ldots \ell_n j m_0\ldots m_n}(\hat{q})\,.
\label{Ptracecond}
\ee

\item It satisfies the transverse condition:
\be
\hat{q}^i  \left( P_{i\ell_0\ell_1 \ldots \ell_n jm_0\ldots m_n}(\hat{q}) + P_{\ell_0 i\ell_1 \ldots \ell_n jm_0\ldots m_n}(\hat{q}) - \frac{2}{3}\delta_{i\ell_0} P_{\ell\ell\ell_1 \ldots \ell_n jm_0\ldots m_n}(\hat{q})\right) = 0\,.
\label{Ptrans}
\ee
\end{enumerate}
See~\cite{Hinterbichler:2013dpa} for a systematic construction and explicit expressions of this operator for
the first few values of $n$. 

We are now in the position to show how the consistency relations~(\ref{wardfinaln>2}) follow from our approach. Substituting~(\ref{correl}) and~(\ref{ambig}), the identity~(\ref{soln}) implies
\beq
\nonumber
& & \frac{\langle \gamma^{jm_0}_{\vec{q}}\zeta_{\vec{p}} \zeta_{-\vec{q} - \vec{p}} \rangle'}{P_\gamma(q)} + \frac{\delta^{jm_0}}{3}\frac{\langle \zeta_{\vec{q}} \zeta_{\vec{p}} \zeta_{-\vec{q} - \vec{p}} \rangle' }{P_\zeta(q)} =   \frac{1}{2}\bigg(\hat{P}^{jm_0k\ell}(\hat{q}) + \frac{2}{3} \delta^{jm_0}\delta^{k\ell}\bigg) K_{k\ell} \\
& &+ \frac{1}{2}\bigg( \hat{P}^{jm_0k\ell}(\hat{q}) + \frac{2}{3} \delta^{jm_0} \delta^{k\ell}(\hat{q}) \bigg) \epsilon_{kcm} \epsilon_{\ell d n}q^c q^d\Big(a( p,q )\delta^{mn} + b( p,q )  p^m p^n\Big)\,.
\label{beforeproject}
\eeq
We are then instructed to differentiate this expression $n$ times with respect to $q$, and project the result using $P_{i\ell_0\ldots \ell_n  j m_0 \ldots m_n}(\hat{q})$. In doing so, we use the
following identities, which, as shown in Appendix C, follow from the properties of the projector:
\beq
\nonumber
& & P_{i \ell_0\ldots \ell_n j m_0\ldots m_n}(\hat{q})\bigg(\hat{P}^{jm_0k\ell}(\hat{q}) + \frac{2}{3} \delta^{jm_0}\delta^{k\ell}\bigg) =2 P_{i \ell_0\ldots \ell_n (k\ell) m_1\ldots m_n}(\hat{q})  \,; \\
\nonumber
& & P_{i \ell_0\ldots \ell_n jm_0\ldots m_n}(\hat{q})\frac{\partial^n \hat{P}_{jm_0 k\ell}(\hat{q})}{\partial q_{m_1}\ldots \p q_{m_n}}=0\,; \\
& & P_{i\ell_0 \ldots \ell_n j m_0\ldots m_n}(\hat{q})\frac{\p^n}{\partial q_{m_1} \ldots \p q_{m_n}} \bigg(  \epsilon^j_{~cm} \epsilon^{m_0}_{~~d n}q^c q^d\Big(a( p,q )\delta^{mn} + b( p,q )  p^m p^n\Big) \bigg)=0\,. 
\label{pr2}
\eeq
It follows that the model-dependent contributions to the identity, encoded in the  last line of~(\ref{beforeproject}), are completely projected out of the consistency relations, as desired. 
Morever, it also follows that all $q$-derivatives go through and hit the $K$ term on the right-hand side of~(\ref{beforeproject}). The result is
\beq
\nonumber
& &  \lim_{\vec{q}\rightarrow 0} P_{i\ell_0 \ldots \ell_n j m_0\ldots m_n}(\hat{q}) \frac{\partial^{n}}{\partial q_{m_1}\cdots \partial q_{m_n}} \Bigg(\frac{1}{P_\gamma(q)} \langle \gamma^{jm_0}_{\vec{q}}\zeta_{\vec{p}} \zeta_{-\vec{q} - \vec{p}}  \rangle' + \frac{\delta^{jm_0}}{3P_\zeta(q)} \langle  \zeta_{\vec{q}}\zeta_{\vec{p}} \zeta_{-\vec{q} - \vec{p}}  \rangle'  \Bigg) \\
&& =  - \lim_{\vec{q}\rightarrow 0} P_{i\ell_0 \ldots \ell_n j m_0\ldots m_n}(\hat{q})  \frac{\partial^{n}}{\partial q_{m_1}\cdots \partial q_{m_n}}  K_{jm_0}\,.
\label{afterproject}
\eeq
Finally, using the rotational invariance of the power spectrum, it is straightforward to show that
\be
P_{i \ell_0\ldots \ell_n j m_0\ldots m_n}(\hat{q}) \bigg(p^{m_0} \frac{\p^{n+1}}{\p p_j \p p_{m_1}\cdots \p p_{m_n}} - p^{m_1} \frac{\p^{n+1}}{\p p_j \p p_{m_0}\p p_{m_2} \cdots \p p_{m_n}} \bigg) P_\zeta( p ) = 0\,.
\ee
It follows that the last line in the expression~(\ref{Kdef}) for $K_{ij}$ projects out. Using this fact,~(\ref{afterproject}) clearly reduces to~(\ref{wardfinaln>2}), as claimed.

The fixed-time path integral method~\cite{Goldberger:2013rsa} used above, while elegant and transparent, has certain limitations. Although the generating functional $W$ derived from~\eqref{3d}
determines correctly the various correlation functions, the vertex functional $\Gamma$ obtained through its Legendre transform is not related straightforwardly to the actual, four-dimensional
effective action.\footnote{In practice, the various contributions to $\Gamma$ are obtained by calculating in-in correlation functions from contact diagrams, and dividing by the power spectrum
for each external leg.} On the other hand, the analyticity assumption made in Sec.~\ref{analytic}, which was critical in deriving the consistency relations, is only well-motivated for
the four-dimensional effective action, while the momentum dependence of (the fixed-time) $\Gamma$ is {\it a priori} unknown. 

To avoid any guesswork, below we will repeat the calculation for the usual time-dependent in-in path integral. In that case the vertex functional encodes effective interaction vertices, which could be read off the quantum action. To leading order in $\hbar$ ({\it i.e.}, at tree level), $\Gamma$ simply encodes the interaction vertices of the classical action we started with, {\it i.e.} GR + inflaton. In this limit the analyticity assumptions about $\Gamma$ directly correspond to assumptions about the locality of the Lagrangian of the theory, and hence the arguments of Sec.~\ref{analytic} about single-field and constant growing modes are well-motivated.

\section{Time-Dependent Path Integral Formalism}
\label{full4d}

In this Section we show how the general consistency relations follow from spatial diffeomorphism invariance of the conventional, time-dependent path integral formalism.
The starting point is the following four-dimensional path integral 
\beq
Z[T,J]=\int \mathcal{D}h_{ij} \mathcal{D}\f e^{iS[h,\f]+i \int{\rm d}^4x (h_{ij} T^{ij}+\f J)}\,.
\eeq
(Note that we omit the gauge condition, it will be imposed when necessary.)
As before, we assume that the lapse function and shift vectors have been integrated out. An important remark is in order here. Whether the above path-integral describes the generator of in-out or in-in diagrams is determined by the time contour, the integration along which determines the action and the source term. For in-out diagrams, the time contour stretches along the real axis $(-\infty,+\infty)$. For in-in diagrams, it lies on the complex plane $(-\infty+i\epsilon,t)\cup (t,-\infty-i\epsilon)$. This approach, followed here, is equivalent to doubling of the fields. The literature on this subject is vast, {\it e.g.}, see~\cite{Berges} and references therein.

The choice of $\zeta$ gauge breaks the time reparametrization symmetry explicitly. The symmetries of interest are therefore spatial diffeomorphisms. Since 
 we are now considering the four-dimensional path integral, in all generality we allow for time-dependent spatial diffeomorphisms $\xi^j(\vec{x},t)$. Demanding
 that $Z[T,J]$ be invariant under~\eqref{sym} and following similar steps as in Sec.~\ref{fixedt}, we obtain the Slavnov-Taylor identity: 
\beq
2\p_j \left( \frac{1}{6} \delta_{ij}\frac{\delta\Gamma}{\delta\zeta}+\frac{\delta \Gamma}{\delta \gamma_{ij}} \right)= \p_i \zeta \frac{\delta \Gamma}{\delta \zeta} + \ldots \,,
\label{tward1}
\eeq
where we have omitted the gauge-fixing contribution for simplicity.
Although this is superficially identical to~\eqref{ward} in the fixed-time approach, an important distinction is that $\Gamma$ now represents the effective action, 
rather than a quantity defined in terms of the equal-time Green's functions. Varying this identity a number of times with respect to $\zeta$ and $\gamma$ leads to various consistency relations
among the vertices of the theory.

As in Sec.~\ref{twohard}, we illustrate this with the simplest case of two hard scalar modes coupled to a soft $\zeta$ or $\gamma$ mode.
Varying~\eqref{tward1} with respect to $\zeta(\vec{x}_1,\tau_1)$ and $\zeta(\vec{x}_2,\tau_2)$, and going to momentum space for the
spatial dimensions, we obtain the master identity
\be
q^j\tilde{\Gamma}_{ij}(\vec{q},\tau;\vec{p},\tau_1;-\vec{q}-\vec{p},\tau_2) = -\delta(\tau-\tau_1)p_i \Gamma_\zeta(|\vec{q}+\vec{p}|; \tau,\tau_2) +  \delta(\tau -\tau_2)(q_i+p_i)\Gamma_\zeta( p;\tau,\tau_1 )\,,
\label{timeward}
\ee
where $\tilde{\Gamma}_{ij}\equiv \frac{1}{3}\delta_{ij}\Gamma^{\zeta\zeta\zeta} + 2\Gamma_{ij}^{\gamma\zeta\zeta}$. This can be translated to a statement about correlation functions using 
\beq
\nonumber
\delta(t_1-t_2)&=&  \int {\rm d} \tau P_\zeta (p;t_1,\tau) \Gamma_\zeta (p;t_2,\tau)\,; \\
\nonumber
\langle \zeta_{\vec{q}}(t)\zeta_{\vec{p}}(t_1)\zeta_{-\vec{q}-\vec{p}}(t_2) \rangle ' &=&\nonumber -\int {\rm d} \tau {\rm d} \tau_1 {\rm d} \tau_2 P_\zeta(q;\tau, t) P_\zeta(p;\tau_1,t_1)P_\zeta(|\vec{q}+\vec{p}|;\tau_2,t_2)\Gamma^{\zeta\zeta\zeta}(\vec{q},\tau;\vec{p},\tau_1;-\vec{q}-\vec{p},\tau_2)\,,\\
\eeq
and similarly for tensors. These relations tell us that we should contract~\eqref{timeward} with two power spectra to obtain:
\beq
q^j\int {\rm d} \tau_1 {\rm d} \tau_2 P_\zeta(p; t ,\tau_1)P_\zeta(|\vec{q}+\vec{p}|; t ,\tau_2)\tilde{\Gamma}_{ij}(\vec{q},\tau;\vec{p},\tau_1;-\vec{q}-\vec{p},\tau_2)=\nonumber \\
\;\;\;\;\;\;\;\;\;\;\;\;\;\;\;\;\;\;\;\;\;\;\;\;\;\;\;\; \delta(\tau-t)\bigg((q_i + p_i) P_\zeta(|\vec{q}+\vec{p}|,\tau) - p_i P_\zeta(p,\tau)\bigg)\,.
\label{tward}
\eeq
As before, this can be solved as a Taylor series around $q = 0$:
\be
\int {\rm d} \tau_1 {\rm d} \tau_2 P_\zeta(p; t ,\tau_1)P_\zeta(|\vec{q}+\vec{p}|; t ,\tau_2)\tilde{\Gamma}_{ij}(\vec{q},\tau;\vec{p},\tau_1;-\vec{q}-\vec{p},\tau_2)= \delta(\tau-t)K_{ij} + A_{ij}(p,q,\tau,t)\,,
\label{tsoln}
\ee
where $K_{ij}$ is given by~(\ref{Kdef}), with $P_\zeta( p )$ understood as $P_\zeta(p,t)$, and $A_{ij}$ again denotes an arbitrary, symmetric and transverse tensor.
(Note that, unlike in Sec.~\ref{twohard}, $A_{ij}$ can now depend on two times.) The scalar and tensor vertices are then isolated by taking the trace and traceless parts of~(\ref{tsoln}).
To extract correlation functions, we multiply the results by the appropriate (unequal-time) power spectra $P(q,t,\tau)$, and integrate over $\tau$:
\beq
\nonumber
\frac{\langle \zeta_{\vec{q}} \zeta_{\vec{p}} \zeta_{-\vec{q} - \vec{p}} \rangle' }{P_\zeta(q)}&=& K+ \int {\rm d}\tau \frac{P_\zeta(q,\tau,t)}{P_\zeta(q,t)} A(p,q,\tau,t) \,;\\
\frac{\langle \gamma^{ij}_{\vec{q}}\zeta_{\vec{p}} \zeta_{-\vec{q} - \vec{p}} \rangle'}{P_\gamma(q)}&=&\frac{1}{2}\hat{P}^{ijk\ell }(\hat{q} )\left( K_{k\ell}+ \int {\rm d}\tau \frac{P_\zeta(q,\tau,t)}{P_\zeta(q,t)}  A_{k\ell}(p,q,\tau,t)\right) \,.
\label{tconv}
\eeq

To derive consistency relations, recall that we had to assume in Sec.~\ref{analytic} that the physical term $A_{ij}$ started at order $q^2$, which was motivated by locality.
A subtlety, already mentioned at the end of Sec.~\ref{twohard}, is that the 3d vertex functional $\Gamma^{3{\rm d}}$ considered in the fixed-time path integral formalism is of course not the same as the 
4d vertex functional $\Gamma^{4{\rm d}}$ of this Section. In particular, their analyticity properties may in principle differ. The meaning of locality for $\Gamma^{4{\rm d}}$ is clear --- it represents the 4d effective action,
which at tree-level reduces to the action we started with (so-called `fundamental action'). Its locality is guaranteed by the locality of the starting-point Lagrangian. Meanwhile, the 3d vertex functional is given by
\beq
\Gamma^{3{\rm d}}(\vec{q},\vec{p},-\vec{p}-\vec{q})\sim \int {\rm d}\tau {\rm d}\tau_1{\rm d}\tau_2 \frac{P_\zeta(q,t,\tau)}{P_\zeta(q,t)} \frac{P_\zeta(p,t,\tau_1)}{P_\zeta(p,t)} \frac{P_\zeta(|\vec{p}+\vec{q}|,t,\tau_2)}{P_\zeta(|\vec{p}+\vec{q}|,t)}\Gamma^{4{\rm d}}(\vec{q},\tau ;\vec{p},\tau_1;-\vec{p}-\vec{q} ,\tau_2)\,.
\label{convert}
\eeq
We see that the locality of $\Gamma^{3{\rm d}}$, on the other hand, follows from that of $\Gamma^{4{\rm d}}$ {\it provided} that the ratio of power spectra $P_\zeta(q,\tau,t)/P_\zeta(q,t)$ is analytic in $q$. This will be the case if mode functions have constant growing-mode solutions. This additional assumption was implicitly made in Sec.~\ref{twohard}. It also implies, incidentally, if $A_{ij}$ starts at order $q^2$, then the time integrals in~\eqref{tconv} will also start at order $q^2$. The rest of the derivation proceeds as in Sec.~\ref{twohard}, and we recover the consistency relations to all orders in $q$.

\section{Conclusion}
\label{conclu}

In this paper, we have shown that the infinite network of consistency relations for adiabatic modes, of which Maldacena's relation is the simplest, all follow from a single
master identity resulting from the Slavnov-Taylor identity for spatial diffeomorphisms. The master identity is cast in terms of the vertex functional and holds for
any momenta. By varying this identity a number of times with respect to the fields, one can obtain consistency relations for the various correlation functions. We have
illustrated this for the simplest case of two hard scalar modes coupled to a soft scalar or tensor mode.

One of the key insights of this derivation is that it makes precise the assumption underlying the consistency relations, namely the locality of the
effective action in the $q\rightarrow 0$ limit. For the simplest inflationary models, this is equivalent to the standard assumption that mode functions tend
to a constant at late times. For more exotic models, in which modes do not ``freeze'' in the usual sense, locality offers an unambiguous criterion. 

The general formalism described here can be applied more broadly to a host other contexts. It should be straightforward to generalize the derivation to include
additional scalar fields. As is well known, consistency relations can be violated in the multi-field context, and it would be interesting to see how this shows up
in our formalism. Other interesting applications include the path integral derivation of consistency relations for the large scale structure~\cite{Kehagias:2013yd,Peloso:2013zw,Creminelli:2013mca,Lamupcoming}, the study of modified initial states~\cite{Holman:2007na,Meerburg:2009ys,Meerburg:2009fi,Ganc:2011dy,Chialva:2011hc,Agarwal:2012mq,Flauger:2013hra,Ashoorioon:2013eia,Aravind:2013lra}, and higher soft limits~\cite{Austinupcoming}.

\vspace{1.5 in}

{\bf Acknowledgements:} We would like to thank Yi-Zen Chu, Paolo Creminelli, Kurt Hinterbichler, Lam Hui, Austin Joyce, Guilherme Pimentel, Marko Simonovic and Junpu Wang for useful discussions. L.B. is supported by 
funds provided by the University of Pennsylvania. J.K. is supported in part by NSF CAREER Award PHY-1145525. While this paper was in its final stages, we became aware of~\cite{Pimentel:2013gza}, which has some overlap
with the results presented here.

\vspace{0.5in}

\section*{Appendix A}
\renewcommand{\theequation}{A-\Roman{equation}}
\setcounter{equation}{0} 

In this Appendix, we show that the projectors $P_{\mu \ell_1\cdots \ell_n \nu m_1 \ldots m_n}$ we have defined for QED (see \eqref{qedp1} and \eqref{qedp2}) are sufficient to project out the model-dependent contribution $C_\mu$ from~\eqref{qedlim}.

First, to ensure that the $q\rightarrow 0$ limit in \eqref{qedlim} is well defined, we need the following identity to hold
\beq
P_{\mu\ell_1\ldots \ell_n \nu m_1 \ldots m_n}\frac{\p^n P^{\nu \alpha}(\hat{q})}{\p q_{m_1}\ldots \p q_{m_n}}=0\,,
\label{pr}
\eeq
for otherwise the derivatives of $P^{\nu \alpha}(\hat{q})$ would yield singular terms as $q\rightarrow 0$. 
Equation~(\ref{pr}) is satisfied, thanks to the properties~\eqref{qedp1} and \eqref{qedp2} of the projector: the derivatives of $P^{\nu\alpha}$ either trace  $P_{\mu\ell_1\ldots \ell_n \nu m_1 \ldots m_n}$
or project it on $q^{m}$.

The contribution of $C_\mu$-dependent terms to the consistency relation is of the form:
\beq
P_{\gamma\ell_1\ldots \ell_n \nu m_1 \ldots m_n}\frac{\p^n }{\p q_{m_1}\ldots \p q_{m_n}}\left( P^{\nu\mu}[(q^2\eta_{\mu\alpha}-q_\mu q_\alpha)v^\alpha+q^\alpha M_{\mu\alpha}] \right)=\nonumber \\
P_{\gamma\ell_1\ldots \ell_n \nu m_1 \ldots m_n}\frac{\p^n }{\p q_{m_1}\ldots \p q_{m_n}}\left(q^2P_{\nu\alpha}v^\alpha+q^\alpha M_{\nu\alpha} \right)\,,
\label{ambigappendix}
\eeq
where in the last step we have used the transversality of $P\mn$ and anti-symmetry of $M\mn$. Using~(\ref{pr}), and noting that $v^\alpha$ and $M\mn$ are both regular in $q\rightarrow 0$ limit, it follows that the $q^2$ and $q^\alpha$ factors in~\eqref{ambig} must be necessarily differentiated. However, differentiating these factors either results in tracing $P_{\gamma\ell_1\ldots \ell_n \nu m_1 \ldots m_n}$, which vanishes by~(\ref{qedp1}), or contracting it with $M\mn$, which vanishes by symmetry. This shows that the properties of $P_{\mu \ell_1\cdots \ell_n \nu m_1 \ldots m_n}$ are sufficient to project out the model-dependent contributions, as claimed.

\section*{Appendix B}
\renewcommand{\theequation}{B-\Roman{equation}}
\setcounter{equation}{0} 

In this Appendix we will justify that, at the tree-level, it is legitimate to neglect the gauge-fixing term, as claimed in Sec.~\ref{fixedt}. In order to keep the consideration simple, 
we will be schematic and omit indices. Focusing on the metric degree of freedom $h$, the generating functional is
\beq
Z[T]=\int \mathcal{D}h ~e^{iS[h]+\int \left( \frac{1}{2\alpha}(\p h)^2 + hT \right)}\,,
\label{toy}
\eeq
where $T$ is an external current, and $\alpha$ is a gauge fixing parameter. We assume that the action, as well as the measure, are invariant under the following schematic diffeomorphism:
\beq
h\rightarrow h+\p\xi+\xi \p h\,.
\label{tr}
\eeq
The invariance of~\ref{toy} under this transformation leads to the following functional differential equation for the generating functional:
\be
\left( \frac{1}{\alpha}\p^3 \frac{\delta}{\delta T(x)}-\p T(x)+T(x) \p \frac{\delta}{\delta T(x)} \right)Z[T]=\frac{1}{\alpha}\int \mathcal{D}h e^{iS[h]+\int \left( \frac{1}{2\alpha}(\p h)^2 + hT \right)}\p^2 h(x) \p h(x)\,.
\label{seq}
\ee
The right hand side is the expectation value of the fields evaluated at the same point. In order to extract the desired relations, we first set $Z=\exp(iW)$ and introduce the effective action $\Gamma[h]$ as a Legende transform of $W$. 
The resulting equation for $\Gamma$ is
\be
\frac{1}{\alpha}\p^3 h(x)-\p \frac{\delta\Gamma}{\delta h(x)}+\p h(x)\frac{\delta\Gamma}{\delta h(x)}= e^{-iW}\frac{1}{\alpha}\int \mathcal{D}h e^{iS[h]+\int \left( \frac{1}{2\alpha}(\p h)^2 + hT \right)} \p^2 h(x) \p h(x)\,.
\label{seq1}
\ee

The first term on the left-hand side is analogous to the first term of~\eqref{qedgamma} --- since we must differentiate this equation at least twice with respect to $h$, this term will not contribute the final identities (similarly to QED).
The gauge-fixing contribution on the right-hand side, on the other hand, is not removable. It is divergent and requires regularization. This is one of the complications associated with non-Abelian gauge theories, compared to Abelian ones. Fortunately, this troublesome term can be ignored at tree level --- it corresponds to a contribution to the vertex functional where fields are evaluated at the same point, and hence is of loop order.

To summarize, at the tree level the vertex functional satisfies
\beq
\frac{1}{\alpha}\p^3 h(x)-\p \frac{\delta\Gamma}{\delta h(x)}+\p h(x)\frac{\delta\Gamma}{\delta h(x)}=0\,.
\eeq
As already mentioned the first term does not contribute to consistency relations. The equation given above is simply a statement about the gauge invariance of the action\footnote{At tree-level, $\Gamma$ coincides with the action $S$, supplemented by the gauge-fixing term.}. In other words, at tree-level there is no need to fix the gauge in vertices. All gauge redundancies are taken care of by the gauge-fixed propagators upon contraction with vertices.

\section*{Appendix C}
\renewcommand{\theequation}{C-\Roman{equation}}
\setcounter{equation}{0} 

In this Appendix we derive identities for $P_{i\ell_0\ldots \ell_n  j m_0 \ldots m_n}$ that are useful in deriving consistency relations. Using the properties listed in Sec.~\ref{high} and the explicit form of $\hat{P}_{j m_0 k \ell}$ from~\eqref{4proj}, it is straightforward to show that
\beq
P_{i \ell_0\ldots \ell_n j m_0\ldots m_n}(\hat{q})\hat{P}_{jm_0 k\ell}(\hat{q})=P_{i \ell_0\ldots \ell_n k\ell\ldots m_n}(\hat{q})+P_{i \ell_0\ldots \ell_n \ell k\ldots m_n(\hat{q})}-\frac{2}{3} \delta_{k\ell} P_{i \ell_0 \ldots \ell_n jj m_1\ldots m_n}(\hat{q})\,.
\label{p1}
\eeq
Note that the structure on the right-hand side is such that it vanishes when hit by $q_k$, which follows from~(\ref{Ptrans}). By tracing~\eqref{p1} we discover another important property of projectors
\beq
P_{i \ell_0\ldots \ell_n j m_0m_1\ldots m_n}(\hat{q})\hat{P}_{jm_0 m_1\ell}(\hat{q})=0\,.
\label{pint}
\eeq

In order to obtain the identities involving derivatives of $\hat{P}_{j m_0 k \ell}$, we will need the fact that its first derivative can be written as
\beq
\frac{\p \hat{P}_{jm_0k\ell}(\hat{q})}{\p q_{m_1}}=-\frac{1}{q^2} \left( q_j \hat{P}_{k \ell m_0 m_1}(\hat{q})+q_k \hat{P}_{j m_0 \ell m_1}(\hat{q})+q_\ell \hat{P}_{j m_0 k m_1}(\hat{q})+q_{m_0} \hat{P}_{k \ell j m_1} (\hat{q})\right)\,.
\label{p2}
\eeq
Contracting this with $P_{i \ell_0\ldots \ell_n j m_0\ldots m_n}$ and using \eqref{p1}, we get
\beq
P_{i \ell_0\ldots \ell_n j m_0\ldots m_n}(\hat{q})\frac{\p \hat{P}_{jm_0k\ell}(\hat{q})}{\p q_{m_1}}=0\,.
\eeq
Having obtained these basic properties of the projector, we proceed by method of strong induction to show that
\beq
P_{i \ell_0\ldots \ell_n jm_0\ldots m_n}(\hat{q})\frac{\partial^n \hat{P}_{jm_0 k\ell}(\hat{q})}{\partial q_{m_1}\ldots \p q_{m_n}}=0\,; \qquad \forall n>1\,,
\label{ans}
\eeq
assuming
\beq
P_{i \ell_0\ldots \ell_n jm_0\ldots m_{n-i}}\frac{\partial^{n-i} \hat{P}_{jm_0 k\ell}(\hat{q})}{\partial q_{m_1}\ldots \p q_{m_{n-i}}}=0\,; \qquad 1\leq i < n\,.
\label{as}
\eeq
Taking into account \eqref{p2}, we have
\beq
&&P_{i \ell_0\ldots \ell_n jm_0\ldots m_n}\frac{\partial^n \hat{P}_{jm_0 k\ell}(\hat{q})}{\partial q_{m_1}\ldots \p q_{m_n}}=P_{i \ell_0\ldots \ell_n jm_0\ldots m_n}\times \nonumber \\
&&\frac{\partial^{n-1}}{\partial q_{m_2}\ldots \p q_{m_n}}\left( -\frac{1}{q^2} \left[ q_j \hat{P}_{k \ell m_0 m_1}+q_k \hat{P}_{j m_0 \ell m_1}+q_\ell \hat{P}_{j m_0 k m_1}+q_{m_0} \hat{P}_{k \ell j m_1} \right] \right)\,.
\label{p3}
\eeq
Upon performing all the differentiations on the right-hand side, we will obtain terms with differentiated $\hat{P}_{jm_0 k\ell}$ as well as with undifferentiated ones. According to the properties of the projector given in Sec.~\ref{high}, along with~\eqref{as}, the only nonzero terms among those involving derivatives of $\hat{P}_{jm_0 k\ell}$ come from differentiating the first term on the right-hand side of \eqref{p3}. Furthermore, among these there are terms with differentiated $q_j$. However, the differentiation of $q_j$ gives us a factor of $\delta_{jm_2}$; as a result, the trace property of $P_{i \ell_0\ldots \ell_n j m_0\ldots m_n}$ becomes applicable and terms under the consideration are nullified by means of the assumption~\eqref{as}.

In other words, the only terms with derivatives of $\hat{P}_{jm_0 k\ell}$ contributing to \eqref{p3} are of the form\footnote{Here, every term in the sum implicitly includes an appropriate multiplicity factor, which we omit for simplicity.}
\beq
P_{i \ell_0\ldots \ell_n jm_0\ldots m_n}\frac{\partial^n \hat{P}_{jm_0 k\ell}(\hat{q})}{\partial q_{m_1}\ldots \p q_{m_n}}\supset &&P_{i \ell_0\ldots \ell_n jm_0\ldots m_n}q_j\times \nonumber \\
&&\sum_{a>0} \frac{\p^{n-a-1}}{\p q_{m_{a+2}}\ldots \p q_{m_{n-1}}}\left(\frac{-1}{q^2} \right) \times \frac{\partial^a \hat{P}_{k\ell m_0 m_1}(\hat{q})}{\partial q_{m_2}\ldots \p q_{m_{a+1}}} \,.
\label{p4}
\eeq
Using the transversality property $P_{i \ell_0\ldots \ell_n jm_0\ldots m_n}q_j=-P_{i \ell_0\ldots \ell_n m_0 j\ldots m_n}q_j+\frac{2}{3} q_{m_0} P_{i \ell_0\ldots \ell_n j j\ldots m_n}$, \eqref{p4} reduces to\footnote{We have used \eqref{as} here again.}
\beq
P_{i \ell_0\ldots \ell_n jm_0\ldots m_n}\frac{\partial^n \hat{P}_{jm_0 k\ell}(\hat{q})}{\partial q_{m_1}\ldots \p q_{m_n}}\supset &&\frac{2}{3}\sum_{a>0} \frac{\p^{n-a-1}}{\p q_{m_{a+2}}\ldots \p q_{m_{n-1}}}\left(\frac{-1}{q^2} \right)\times \nonumber \\
&& P_{i \ell_0\ldots \ell_n jjm_1\ldots m_n}q_{m_0} \frac{\partial^a \hat{P}_{k\ell m_0 m_1}(\hat{q})}{\partial q_{m_2}\ldots \p q_{m_{a+1}}}\,.
\eeq
Now, using the expression for $a$th derivative of the identity $q_{m_0}\hat{P}_{k\ell m_0 m_1}=0$, it is easy to see that the only surviving term in the sum will be the one with $a=1$. Furthermore, there will be $n-1$ of those terms in \eqref{p3}. Hence, the only term involving the derivative $\hat{P}_{jm_0k\ell}$, using $q_{m_0}\p_{q_{m_2}}\hat{P}_{k\ell m_0 m_1}=-\hat{P}_{k\ell m_2 m_1}$, reduces to
\beq
P_{i \ell_0\ldots \ell_n jm_0\ldots m_n}\frac{\partial^n \hat{P}_{jm_0 k\ell}(\hat{q})}{\partial q_{m_1}\ldots \p q_{m_n}}\supset -\frac{2}{3}(n-1) \frac{\p^{n-2}}{\p q_{m_{3}}\ldots \p q_{m_{n-1}}}\left(\frac{-1}{q^2} \right)
P_{i \ell_0\ldots \ell_n jjm_1\ldots m_n} \hat{P}_{k\ell m_1 m_2}(\hat{q})\,.
\label{f1}
\eeq
The rest of the terms in \eqref{p3} are the ones with no derivative acting on $\hat{P}_{jm_0 k\ell}$ to begin with. The only non-vanishing ones are with one derivative acting on $q_{i}$
\beq
P_{i \ell_0\ldots \ell_n jm_0\ldots m_n}\frac{\partial^n \hat{P}_{jm_0 k\ell}(\hat{q})}{\partial q_{m_1}\ldots \p q_{m_n}}\supset (n-1)  \frac{\p^{n-2}}{\p q_{m_{3}}\ldots \p q_{m_{n-1}}}\left(\frac{-1}{q^2} \right) \times \nonumber \\
P_{i \ell_0\ldots \ell_n jm_0\ldots m_n} \left[ \delta_{jm_2} \hat{P}_{k \ell m_0 m_1}+\delta_{km_2} \hat{P}_{j m_0 \ell m_1}+\delta_{\ell m_2} \hat{P}_{j m_0 k m_1}+\delta_{m_0m_2} \hat{P}_{k \ell j m_1} \right]\,.
\label{f2}
\eeq
Combining~\eqref{f1} and~\eqref{f2}, we obtain
\beq
P_{i \ell_0\ldots \ell_n jm_0\ldots m_n}\frac{\partial^n \hat{P}_{jm_0 k\ell}(\hat{q})}{\partial q_{m_1}\ldots \p q_{m_n}}=(n-1)  \frac{\p^{n-2}}{\p q_{m_{3}}\ldots \p q_{m_{n-1}}}\left(\frac{-1}{q^2} \right) \times \nonumber \\
\left(- \frac{2}{3}P_{i \ell_0\ldots \ell_n jjm_1\ldots m_n} \hat{P}_{k\ell m_1 m_2}+P_{i \ell_0\ldots \ell_n jm_0\ldots m_n} \left[ \delta_{jm_2} \hat{P}_{k \ell m_0 m_1}+\delta_{m_0m_2} \hat{P}_{k \ell j m_1} \right]\right)\,,
\eeq
where we have used \eqref{pint}. The right-hand side vanishes, once we use the trace property of the projector. In other words, the identity~(\ref{ans}) holds, as we wanted to show.

\end{document}